\documentstyle[aps,preprint,tighten,epsf]{revtex}

\begin{document}

\preprint{}
\draft

\title{The NMR relaxation rate of $^{17}$O in 
Sr$_2$CuO$_2$Cl$_2$: Probing two-dimensional magnons
at short distances
 }
\author{Peter Kopietz$^{1}$ and Sudip Chakravarty$^{2}$} 
\address{
$^{1}$Institut f\"{u}r Theoretische Physik der Universit\"{a}t G\"{o}ttingen,\\
Bunsenstr.9, D-37073 G\"{o}ttingen, Germany\\
$^2$Department of Physics and Astronomy, University of California, Los Angeles, CA 90095}
\date{February 18, 1997}
\maketitle
\begin{abstract}
We calculate the 
nuclear relaxation rate $1/T^{\rm O}_1 $ of  
oxygen
in the undoped quasi two-dimensional  quantum Heisenberg
antiferromagnet Sr$_2$CuO$_2$Cl$_2$ above the Neel temperature.
The calculation is performed at two-loop order
with the help of the Dyson-Maleev formulation of the spin-wave expansion,
taking all scattering processes
involving two and three magnons into account.
At low temperatures $T$ we find
$1 / T^{\rm O}_1  = c_1 T^3 + c_2 T^4 + O (T^5)$, and give 
explicit expressions for the coefficients
$c_1$ (two-magnon scattering) and $c_2$ (three magnon scattering).
We compare our result with a recent experiment by Thurber {\it{et al.}}
and show that this experiment directly probes the existence 
of short-wavelength magnons in a two-dimensional antiferromagnet.

\end{abstract}
\narrowtext

\section{Introduction}
\label{sec:intro}

In a recent experiment\cite{Thurber96}  Thurber {\it{et al.}} reported
nuclear magnetic resonance (NMR) measurements of the $^{17}$O and $^{63}$Cu relaxation
rates of the undoped and lightly doped quasi two-dimensional quantum Heisenberg
antiferromagnet Sr$_2$CuO$_2$Cl$_2$.
In this compound the  anisotropies and inter-layer couplings are particularly weak, so 
that in the undoped case it is an excellent approximation to model the dynamics of the
localized $S = 1/2$ spins at the Cu sites
above the Neel temperature 
$T_N$ by an isotropic two-dimensional nearest neighbor quantum
Heisenberg antiferromagnet. The Hamiltonian is given by
 \begin{equation}
 \hat{H} = J \sum_{ {\bf{r}} } \sum_{i = x,y}
 {\bf{S}}_{\bf{r}} \cdot {\bf{S}}_{ {\bf{r}}+ {\bf{a}}_i }
 \label{eq:Hamiltonian}
 \; \; \; ,
 \end{equation}
where the ${\bf{r}}$-sum is over the $N$ sites of the square lattice formed by
the Cu-sites, and 
${\bf{a}}_x$ and ${\bf{a}}_y$ are the primitive lattice vectors
with length $| {\bf{a}}_i | = a$.
The operators ${\bf{S}}_{\bf{r}}$ are $SU (2)$ spin operators
satisfying ${\bf{S}}_{\bf{r}}^2 = S ( S + 1)$, and $J > 0$ is the antiferromagnetic
exchange coupling.
NMR experiments measure the relaxation of the nuclear spins
due to their coupling to the reservoir of the
electronic spin lattice.
The theoretical framework for the calculation of the NMR rates
in Heisenberg antiferromagnets
has been discussed 30 years ago by Beemann and Pincus\cite{Beeman68},
and more recently for two-dimensional antiferromagnets in 
Refs.\cite{Mila89,Mila89b,Orbach90,Chacky91}.
However, so far only the leading order of the low-temperature 
behavior has been calculated, which is determined by two-magnon scattering and
can be obtained from a simple one-loop calculation.
For a quantitative comparison with the 
experimental result\cite{Thurber96}
for the NMR rate of $^{17}$O 
it is important to know also the
contribution from three-magnon scattering, which 
describes the leading effect of magnon-magnon interactions.
In this work we shall explicitly calculate this correction and compare it
with the experiment of Thurber {\it{et al.}}\cite{Thurber96}.

\begin{figure}[htb]
\vspace{1.0cm}
\hspace{3.0cm}
\epsfysize6cm
\epsfbox{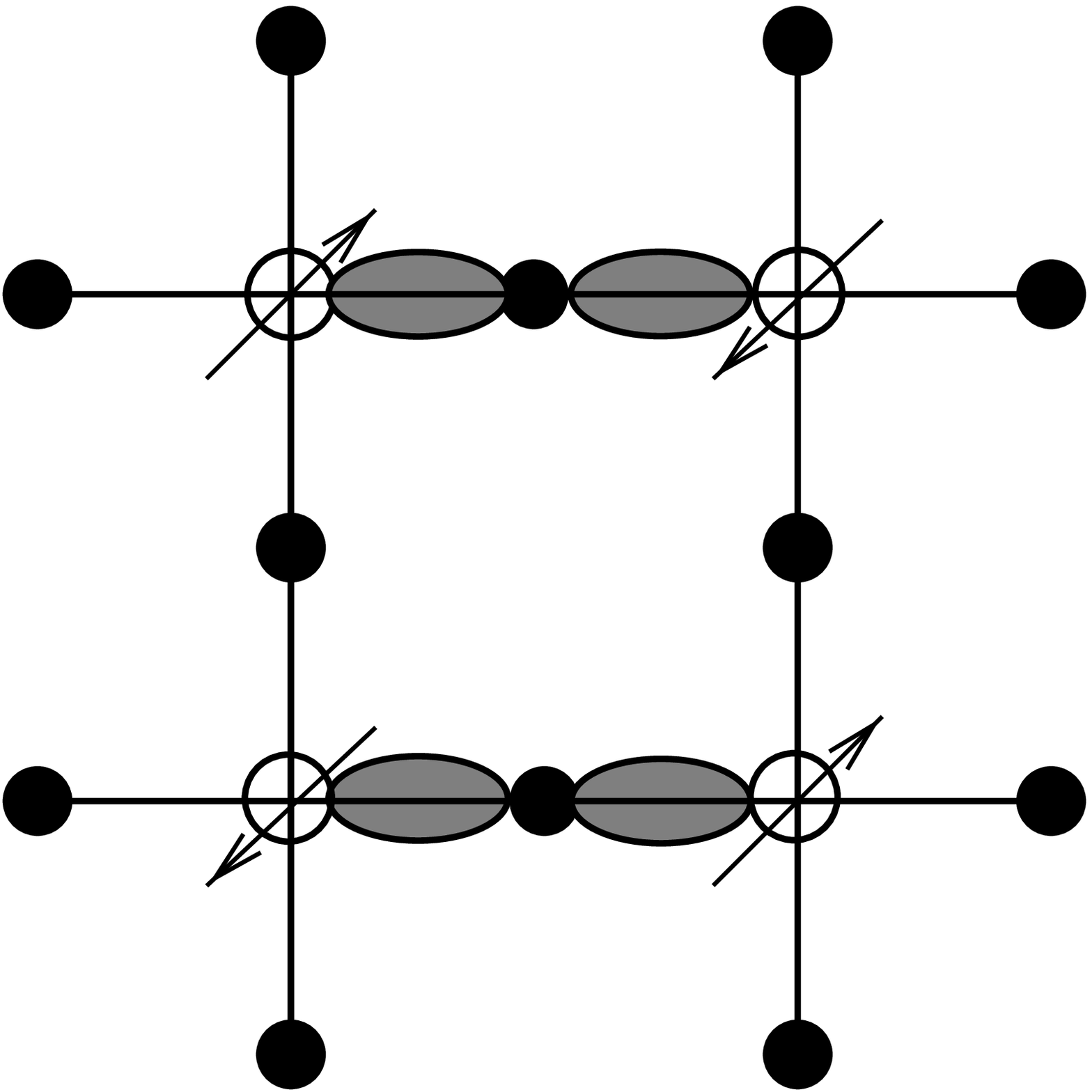}
\caption{ 
Cu-O planes and hyperfine couplings.
The black dots mark the positions of the oxygen atoms, and the
empty circles are the copper atoms with the
electronic spins. The shaded areas symbolize the hyperfine
coupling between a given oxygen site and the  two
neighboring copper spins.
}
\vspace{1.0cm}
\label{fig:plane}
\end{figure}

To begin with, let us
briefly derive the NMR rate from Fermi's golden rule.
Note that Mila and Rice\cite{Mila89b} gave the relevant 
form factor only
up to a constant of proportionality. 
However, for a quantitative comparison with experiments we need to
know the precise numerical value of the form factor.
In the simplest approximation, the coupling between the
nucleus and the electronic spins can be described by
an isotropic hyperfine interaction\cite{Beeman68,Mila89,Mila89b,Orbach90,Chacky91}.
For a nuclear spin ${\bf{I}}^{\rm Cu}_{\bf{R}}$ of $^{63}$Cu at
position ${\bf{R}}$ this is of the form
 $\hat{H}^{\rm Cu } = A_{\rm Cu} 
 {\bf{I}}^{\rm Cu }_{\bf{R}} \cdot {\bf{S}}_{ {\bf{R}}  }$,
with some characteristic energy scale $A_{\rm Cu}$  that can be obtained
from experiments. 
As discussed by Chakravarty and Orbach\cite{Orbach90}, 
the $^{63}$Cu NMR-rate is dominated by critical antiferromagnetic fluctuations,
and cannot be calculated via the perturbative spin-wave expansion.
On the other hand, the nuclear
spins ${\bf{I}}^{\rm O}_{\bf{R}}$ at the $^{17}$O-site are
coupled to the electronic spins of two
neighboring Cu sites (see Fig.~\ref{fig:plane}), so that
the hyperfine interaction is
 \begin{equation}
 \hat{H}^{\rm O } = A_{\rm O} 
 {\bf{I}}^{\rm O }_{\bf{R}}  \cdot \sum_{i = 1,2}  {\bf{S}}_{ {\bf{R}} + {\bf{b}}_i } 
 \label{eq:Ohf}
 \; \; \; ,
 \end{equation}
where ${\bf{b}}_1 = - {\bf{a}}_x / 2$ and ${\bf{b}}_2 = {\bf{a}}_x /2$.
Let us emphasize that Eq.(\ref{eq:Ohf}) involves the
{\it{average}}
 $ {\bf{S}}_{ {\bf{R}} + {\bf{a}}_x /2 }  + 
 {\bf{S}}_{ {\bf{R}} - {\bf{a}}_x /2 }  $ 
of two neighboring electronic spins.
Given the fact that the spins couple 
antiferromagnetically, it is clear that
at least in the classical picture the total spin vanishes.
Hence, we expect that
the $^{17}$O relaxation rate is dominated by short wavelength
antiferromagnetic quantum fluctuations. In this work we shall
put this expectation on a solid quantitative basis,
and show that the NMR relaxation rate $1/T_1^{\rm O}$
at the oxygen site probes the  unique short-distance
behavior of magnons in two-dimensional Heisenberg magnets.

Keeping in mind that $1/T_1$ is defined such that calculations
should always use the matrix elements for nuclear spins
with magnitude $I = 1/2$,
straightforward application of Fermi's golden rule yields 
for the NMR relaxation rate of 
$^{17}$O
 \begin{equation}
 \frac{1}{T_1^{\rm O}} = \frac{ A_{\rm O}^2}{3 \hbar^2} 2 \pi \hbar \sum_{\alpha \beta}
 p_{\alpha} \langle \alpha |  \sum_i {\bf{S}}_{{\bf{R}} + {\bf{b}}_i }  | \beta \rangle \cdot
  \langle \beta |  \sum_{j} 
  {\bf{S}}_{{\bf{R}} + {\bf{b}}_j } | \alpha \rangle
  \delta ( \hbar \omega_N - E_{\alpha} + E_{\beta} )
  \label{eq:NMRTO}
  \; \; \; ,
  \end{equation}
where $E_{\alpha}$ and $ | \alpha \rangle$ are exact eigen-energies
and eigen-states of the Heisenberg Hamiltonian  (\ref{eq:Hamiltonian}), and
$p_{\alpha} = e^{- E_{\alpha} / T } / Z$, where $Z = \sum_{\alpha} e^{- E_{\alpha} / T}$.
Here $T$ is the temperature, measured in units of energy.
In experiments the resonance energy $\hbar \omega_N$ is usually much smaller than
all characteristic energies of the spin system, so that we may set
$\omega_N = 0$.
Defining the Fourier transformed spin operators 
$ {\bf{S}}_{\bf{r}} = {{N}}^{-1/2} \sum_{\bf{k}} e^{-  i {\bf{k}} \cdot {\bf{r}} }
 {\bf{S}}_{\bf{k}}$,
where the ${\bf{k}}$-sum is over the first Brillouin zone of the
reciprocal lattice, we obtain from Eq.(\ref{eq:NMRTO})
for $\omega_{N} \rightarrow 0$
 \begin{equation}
 \frac{1}{T_1^{\rm O}} = 
 \frac{ A_{\rm O}^2}{3 \hbar^2N } \sum_{\bf{k}} f ( {\bf{k}} ) 
 S ( {\bf{k}} , 0 ) 
 \; \; \; ,
 \label{eq:NMRdynstruc}
 \end{equation}
where $S ( {\bf{k}} , \omega )$ is the dynamic structure factor
of the Cu-spin lattice,
 \begin{equation}
 S ( {\bf{k}} , \omega )  =
 2 \pi \hbar \sum_{\alpha \beta}
 p_{\alpha} \langle \alpha |  {\bf{S}}_{\bf{k}}  | \beta \rangle \cdot
  \langle \beta |  {\bf{S}}_{ - \bf{k}} | \alpha \rangle
  \delta ( \hbar \omega - E_{\alpha} + E_{\beta} )
  \; \; \; ,
  \label{eq:dynstruc}
  \end{equation}
and the form factor is\cite{footnote1}
 \begin{equation}
 f ( {\bf{k}} )  =  2 \left[ 1 + \cos ( k_x a ) \right]
 \label{eq:fO}
 \; \; \; .
 \end{equation}
Note that a transition between the two states of the nuclear spin 
is described by the ladder operators $I^{\pm} = I^{x} \pm i I^{y}$, which couple to the 
transverse components $S^{\mp} = S^{x} \mp i S^{y}$ of the
electronic spins. However, for temperatures above the Neel temperature
the system is rotationally invariant in spin space, 
which enables us to express the NMR rate in terms of 
the rotationally invariant dynamic structure factor (\ref{eq:dynstruc}).
This is a very important point, because we would like
to calculate the dynamic structure factor in Eq.(\ref{eq:NMRdynstruc})
by means of the spin-wave expansion. At the first sight
it seems that for two-dimensional
Heisenberg magnets at finite temperatures
this approach cannot be
justified, because the naive spin-wave expansion is based on the assumption
of long-range order, which is rigorously known
to be absent at any finite temperature\cite{Hohenberg67}.
For example, an attempt to calculate the staggered magnetization
at $ T > 0$ via spin-wave theory 
leads to a infrared divergent integral, signalling the
inconsistency of the magnon picture in two dimensions\cite{Chakravarty90}.
The important point is, however, that {\it{spin-rotationally invariant}} quantities
are free of infrared divergencies, and can be calculated with the help
the spin-wave expansion even at finite temperatures.
For the classical Heisenberg model this has been proven by David\cite{David81}, and
it is reasonable to assume that it remains true for the quantum model.
In this work we shall verify explicitly at two-loop order
that in the
rotationally invariant dynamic structure factor (\ref{eq:dynstruc})
all infrared divergencies that appear at intermediate stages of the
calculation indeed cancel. 

\section{The spin-wave expansion for the $^{17}$O NMR rate}
\label{sec:DM}

In this section we shall briefly describe our procedure for  calculating 
the dynamic structure factor with the help of the Dyson-Maleev
formulation of the spin-wave expansion\cite{Dyson56}. 
As compared with the more common Holstein-Primakoff formalism\cite{Holstein40},
the Dyson-Maleev spin-boson mapping has the advantage that 
the boson representation of the Hamiltonian involves only 
one- and two-body terms,
so that beyond the leading order
in the spin-wave expansion the number of
Feynman diagrams generated with the help of the Dyson-Maleev transformation
is smaller than in the case of the Holstein-Primakoff formalism.
Thus, for our two-loop calculation
of the dynamic structure factor the 
Dyson-Maleev transformation leads to considerable
technical simplifications.
Of course, final results for physical quantities should be identical
in both formalisms.  

\begin{figure}[htb]
\vspace{1.0cm}
\hspace{3.0cm}
\epsfysize6cm
\epsfbox{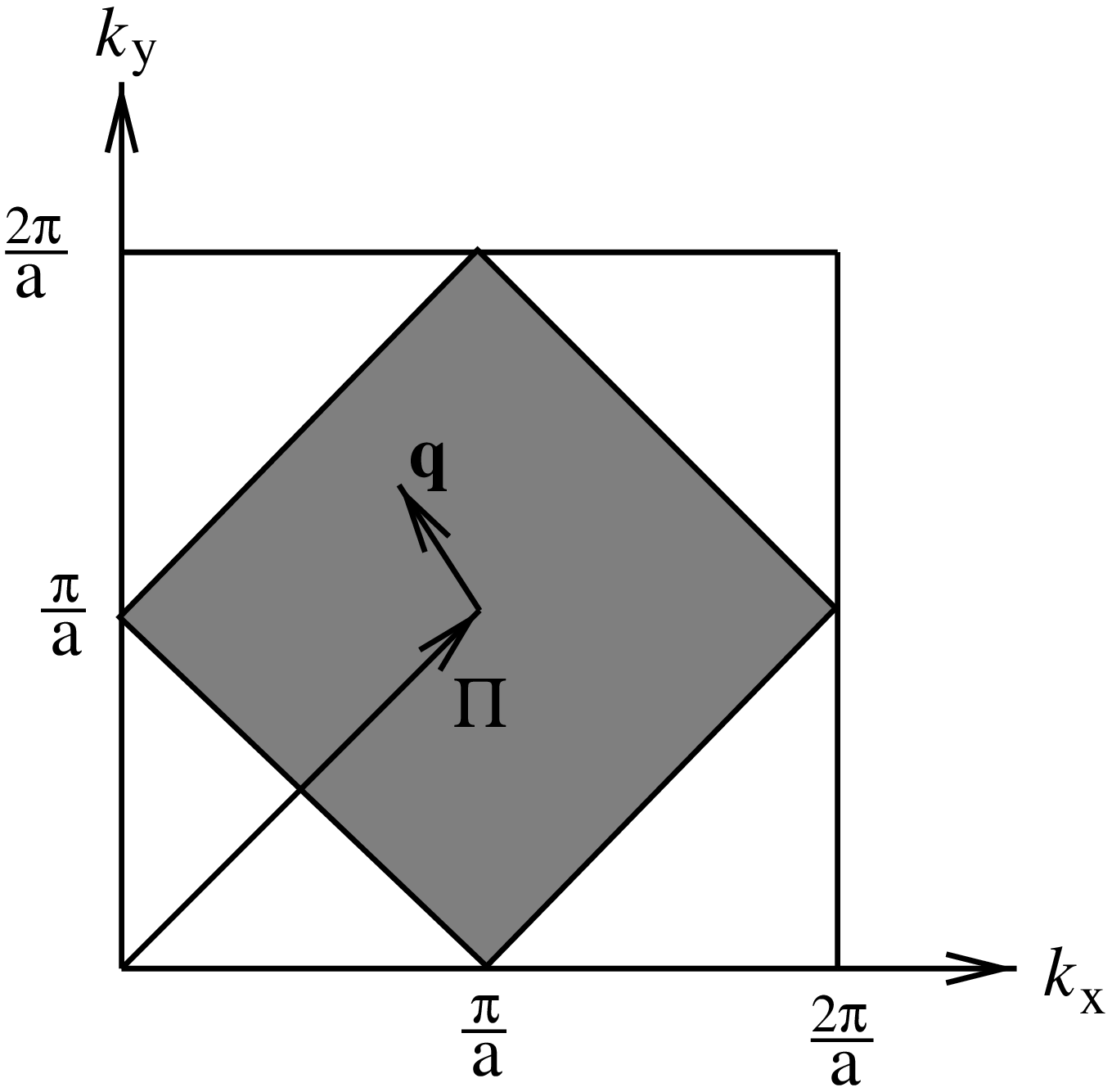}
\caption{ 
First Brillouin zone of a square lattice and reduced
Brillouin zone of the antiferromagnet (shaded area).
In the reduced zone scheme all wave-vectors ${\bf{q}}$ are measured
with respect to the antiferromagnetic ordering wave-vector ${\bf{\Pi}} = ( \pi / a , \pi /a)$.}
\vspace{1.0cm}
\label{fig:BZ}
\end{figure}

As usual, we divide the square lattice into
two sublattices, labelled  $A$ and $B$, such that the nearest neighbors
of any given site belong to the other sublattice.
The 
spherical components of the spin operators
on the $A$-sublattice are then represented by
canonical boson operators $a_{\bf{r}}$ in the following manner,
 \begin{equation}
 \begin{array}{lcl}
 S^{+}_{\bf{r}} & = & \sqrt{2S} \left[ 1 - (2 S)^{-1} a_{\bf{r}}^{\dagger} a_{\bf{r}}
 \right] a_{\bf{r}}
 \\
 S^{-}_{\bf{r}} & = & \sqrt{2S} a^{\dagger}_{\bf{r}} 
 \\
 S^{z}_{\bf{r}} & = & S - a^{\dagger}_{\bf{r}} a_{\bf{r}}
 \end{array}
 \; \; \; , \; \; \; {\bf{r}} \in A
 \;  .
 \label{eq:DMa}
 \end{equation}
Similarly we write on 
the $B$-sublattice
 \begin{equation}
 \begin{array}{lcl}
 S^{+}_{\bf{r}} & = & \sqrt{2S} b^{\dagger}_{\bf{r}} 
 \left[ 1 - (2 S)^{-1} b_{\bf{r}}^{\dagger} b_{\bf{r}}
 \right]
 \\
 S^{-}_{\bf{r}} & = & \sqrt{2S} b_{\bf{r}} 
 \\
 S^{z}_{\bf{r}} & = & - S + b^{\dagger}_{\bf{r}} b_{\bf{r}}
 \end{array}
 \; \; \; , \; \; \; {\bf{r}} \in B
 \;  ,
 \label{eq:DMb}
 \end{equation}
where $b_{\bf{r}}$ are again canonical boson operators.
Because of the two-sublattice structure, it is convenient to
Fourier transform on both sublattices separately,
 \begin{equation}
 a_{\bf{r}}  =  ( N / 2 )^{-1/2} {\sum_{\bf{q}}}^{\prime} e^{i {\bf{q}} \cdot
 {\bf{r}}} a_{\bf{q}} \; \; \; , \; \; \; 
 b_{\bf{r}}  =  ( N / 2 )^{-1/2} {\sum_{\bf{q}}}^{\prime} e^{i {\bf{q}} \cdot
 {\bf{r}}} b_{\bf{q}} 
 \label{eq:FTab}
 \; \; \; ,
 \end{equation}
where the prime
indicates that the wave-vector sums are over
the $N/2$ points of the reduced Brillouin zone, 
with the wave-vectors ${\bf{q}}$ measured with respect to the antiferromagnetic
ordering wave-vector ${\bf{\Pi}} = [ \pi / a , \pi / a ]$ (see 
Fig.\ref{fig:BZ}). 
Because we would like to express all sums 
consistently in terms of the wave-vectors of the reduced Brillouin zone,
we rewrite Eq.(\ref{eq:NMRdynstruc}) as
 \begin{equation}
 \frac{1}{T_1^{\rm O}} = 
 \frac{ A_{\rm X}^2}{3 \hbar^2N } {\sum_{\bf{q}}}^{\prime} \left[ 
 f ( {\bf{q}} ) S ( {\bf{q}} , 0 ) 
 +
 f_{\rm st} ( {\bf{q}} ) 
 {S}_{\rm st} ( {\bf{q}} , 0 ) 
 \right]
 \; \; \; ,
 \label{eq:NMRdynstruc2}
 \end{equation}
 where
  ${S}_{\rm st} ( {\bf{q}} , \omega  )   \equiv 
  S ( {\bf{\Pi}} + {\bf{q}} , \omega  ) $ is the {\it{staggered}}
  structure factor, and  
  $f_{\rm st} ( {\bf{q}} )   \equiv  
  f ( {\bf{\Pi}} + {\bf{q}} ) $ is the corresponding form factor. 
Below it will become obvious that at low temperatures the
inverse thermal de Broglie wavelength
acts as an ultraviolet cutoff for the
${\bf{q}}$-sums in Eq.(\ref{eq:NMRdynstruc2}).  
Note that the thermal de Broglie wavelength 
(divided by $2 \pi $) of an isotropic antiferromagnet can be written as
 \begin{equation}
 \lambda_{\rm th} = \frac{\hbar c }{T}
 \; \; \; ,
 \label{eq:deBrogliedef}
 \end{equation}
where $c$ is the spin-wave velocity.
Hence, at low temperatures $\lambda_{\rm th} \gg a$, so that
the sums in Eq.(\ref{eq:NMRdynstruc2}) 
are dominated by a small circle 
in the ${\bf{q}}$-plane  with radius $\lambda_{\rm th}^{-1}$ 
around the origin.  
Obviously the first term in Eq.(\ref{eq:NMRdynstruc2})
represents the contribution from fluctuations of the
{\it{total magnetization}}. Because
these are not critical fluctuations, we expect
that the corresponding contribution  can be calculated
perturbatively.
From Eq.(\ref{eq:fO}) we see that for small ${\bf{q}}$
the form factor for $1/T_1^{\rm O}$ can be replaced by
 \begin{equation}
 f ( {\bf{q}} )  =  4 + O ( q_x^2 )
 \; \; \; .
 \label{eq:fO1}
 \end{equation}
On the other hand, the second term in Eq.(\ref{eq:NMRdynstruc2})
represents for $| {\bf{q}} | \ll a^{-1}$ long wavelength
critical fluctuations of the staggered magnetization.
However, the signature of the critical fluctuations contained in
  $S_{\rm st} ( {\bf{q}} , 0  )  $ 
is to a large extent removed  by the form factor, which vanishes for small
${\bf{q}}$ as
 \begin{equation}
 f_{\rm st} ( {\bf{q}} )  =  ( q_x a )^2 + O ( q_x^4 )
 \label{eq:fstO1}
 \; \; \; .
 \end{equation}
In fact, we shall show shortly that  the contributions of
both terms on the right-hand side of
Eq.(\ref{eq:NMRdynstruc2}) have the same order
of magnitude,
so that the $^{17}$O relaxation rate is determined both by
non-critical fluctuations of the total magnetization and by
short wavelength antiferromagnetic fluctuations.
In contrast, the corresponding form factor for 
$^{63}$Cu is unity for all wave-vectors\cite{Mila89,Mila89b,Orbach90},
so that the rate $1/T_1^{\rm Cu}$ is completely dominated by
critical antiferromagnetic fluctuations\cite{Orbach90}.

The spin-wave expansion for the
dynamic spin-spin correlation functions has been described
in an impressive paper by  Harris, Kumar, Halperin and Hohenberg\cite{Harris71},
so that we can be rather brief here and refer the reader to
Ref.\cite{Harris71} for more details.
After substituting the Dyson-Maleev transformation 
(\ref{eq:DMa},\ref{eq:DMb})
for the spin operators into Eq.(\ref{eq:Hamiltonian}) and Fourier transforming the
boson operators as in Eq.(\ref{eq:FTab}), the spin Hamiltonian in $D$ dimensions
is mapped onto the following boson Hamiltonian
 \begin{equation}
 \hat{H} \rightarrow - N D J S^2 + \hat{H}_{\rm DM}^{(2)} + \hat{H}_{\rm DM}^{(4)}
 \label{eq:Hmap}
 \; \; \; ,
 \end{equation}
with
 \begin{equation}
 \hat{H}_{\rm DM}^{(2)} = 2 D J S {\sum_{\bf{q}}}^{\prime}
 \left[ a^{\dagger}_{\bf{q}} a_{\bf{q}} + b^{\dagger}_{\bf{q}} b_{\bf{q}}
 + \gamma_{\bf{q}} ( a_{\bf{q}} b_{\bf{q}} + a^{\dagger}_{\bf{q}} b^{\dagger}_{\bf{q}} )
 \right]
 \label{eq:HDM2}
 \; \; \; ,
 \end{equation}
 \begin{equation}
 \hat{H}_{\rm DM}^{(4)} = -  \frac{2 D J }{ N}
 {\sum_{ {\bf{1}} \ldots {\bf{4}} }}^{\prime}
 \delta_{\ast} ( {\bf{3}} + {\bf{4}} - {\bf{1}} - {\bf{2}} )
 \left[ 2 \gamma_{ {\bf{2 -4 }} } 
 a^{\dagger}_{\bf{1}} a_{\bf{3}} b^{\dagger}_{\bf{4}} b_{\bf{2}}  
 + \gamma_{\bf{2}} 
 a^{\dagger}_{\bf{1}} a_{\bf{3}} a_{\bf{4}} b_{\bf{2}}  
 + \gamma_{\bf{3+4-2}} 
 a^{\dagger}_{\bf{1}} b^{\dagger}_{\bf{3}} b^{\dagger}_{\bf{4}} b_{\bf{2}}  
 \right]
 \label{eq:HDM4}
 \; \; \; ,
 \end{equation}
where ${\bf{1}}$ for ${\bf{q}}_1$ (and similarly for the other labels), and
we have introduced the standard notation
 $\gamma_{\bf{q}} = D^{-1} \sum_{i = 1}^{D} \cos ( {\bf{q}} \cdot {\bf{a}}_i )$.
The symbol $\delta_{\ast} ( {\bf{q}} )$ indicates momentum conservation
up to a vector of the reciprocal lattice associated with 
the sublattices\cite{Kopietz93}.
The quadratic part $H_{\rm DM}^{(2)}$ of the Dyson-Maleev Hamiltonian is now
diagonalized by means of a Bogoliubov transformation,
 \begin{equation}
 \left( \begin{array}{c}
 a_{\bf{q}} \\
 b^{\dagger}_{\bf{q}}
 \end{array}
 \right)
 = u_{\bf{q}} \left( \begin{array}{cc}
 1  & - x_{\bf{q}} \\
 - x_{\bf{q}} &  1
 \end{array}
 \right)
 \left( \begin{array}{c}
 \alpha_{\bf{q}} \\
 \beta_{\bf{q}}^{\dagger}
 \end{array}
 \right)
 \label{eq:Bogoliubov}
 \; \; \; ,
 \end{equation}
with 
 \begin{equation}
 u_{\bf{q}}  =  \sqrt{ \frac{1 + \epsilon_{\bf{q}} }{2 \epsilon_{\bf{q}} }}
 \; \; \; , \; \; \; 
 x_{\bf{q}}  =  \sqrt{ \frac{1 - \epsilon_{\bf{q}} }{1 +  \epsilon_{\bf{q}} }}
 \label{eq:xqdef}
 \; \; \; ,
 \end{equation}
where
 $\epsilon_{\bf{q}}  =  \sqrt{ 1 - \gamma_{\bf{q}}^2 }$.
Then we obtain 
 \begin{equation}
 \hat{H}_{\rm DM}^{(2)} = - 2 D J S + {\sum_{\bf{q}}}^{\prime}
 E_{\bf{q}} \left[ \alpha^{\dagger}_{\bf{q}} \alpha_{\bf{q}}
 + \beta_{\bf{q}} \beta^{\dagger}_{\bf{q}}  \right]
 \label{eq:HDM2bogo}
 \; \; \; .
 \end{equation}
Here $E_{\bf{q}}  = 2 D J S \epsilon_{\bf{q}}$ is the free
magnon dispersion.
Taking into account that the $\beta$-operators in Eq.(\ref{eq:HDM2bogo})
are anti-normal ordered, we obtain
the following $1/S$-correction to the ground-state energy,
$\delta E_0 = -2 DJS C$, with $C = 1 - 
( 2/N  ) {\sum_{\bf{q}}}^{\prime} \epsilon_{\bf{q}}$. 
The numerical value of $C$ has first been
calculated by Anderson \cite{Anderson52}. In two dimensions the result is
$C = 0.158$.
In our two-loop calculation presented in Sec.\ref{sec:twoloop}
we shall simply ignore similar zero-temperature renormalizations,
which involve higher powers of $1/S$. 
Quantum corrections of this type are implicitly taken into account
by identifying the spin-wave velocity $c$ and the spin stiffness
$\rho_s$ in our final result (see Eq.(\ref{eq:TOres}) below)
with the experimentally measured values.
Substituting the Bogoliubov transformation (\ref{eq:Bogoliubov}) into
the quartic part $\hat{H}_{\rm DM}^{(4)}$ we obtain totally $10$ different terms describing
scattering of the magnons $\alpha_{\bf{q}}$ and $\beta_{\bf{q}}$
in various combinations. 
The vertices for these scattering processes have been
derived in Ref.\cite{Harris71}, and will not be reproduced here. 
For our two-loop calculation we shall use a particular symmetric
parameterization of the Dyson-Maleev vertices given in
Ref.\cite{Kopietz90}.
It turns out that the complete interaction part of the
Dyson-Maleev Hamiltonian can be
parameterized in terms of five different vertices, 
which can be written as
 \begin{equation}
 V^{(j)} ( {\bf{q}} , {\bf{k}} , {\bf{k}}^{\prime} )
 = \frac{ u_{ {\bf{k}} + {\bf{q}} } u_{  {\bf{k}}^{\prime} - {\bf{q}} } }{u_{\bf{k}}
 u_{ {\bf{k}}^{\prime} } } 
 \tilde{V}^{(j)} ( {\bf{q}} , {\bf{k}} , {\bf{k}}^{\prime} )
 \; \;  ,  \; \; j = 1 , \ldots , 5 
 \; \; ,
 \label{eq:VDMdef}
 \end{equation}
where the rescaled vertices $\tilde{V}^{(j)}$ are non-singular.
For our purpose it is sufficient to know
the leading behavior of these vertices for small 
wave-vectors\cite{Harris71,Kopietz90,Kopietzphd}, 
 \begin{equation}
 \tilde{V}^{(j)} ( {\bf{q}} , {\bf{k}} , {\bf{k}}^{\prime} )
 \sim 
 \left\{
 \begin{array}{ll}
 \frac{1}{2} ( 1 -  \hat{\bf{k}} \cdot \hat{\bf{k}}^{\prime} ) &
 \mbox{for $  j=1,2,5$} \\
 \frac{1}{2} ( 1 +  \hat{\bf{k}} \cdot \hat{\bf{k}}^{\prime} ) &
 \mbox{for $ j=3,4$}
 \end{array}
 \right.
 \; \; \; ,
 \label{eq:VDMasym}
 \end{equation}
where $\hat{\bf{k}} = {\bf{k}} / | {\bf{k}} | $ and
$\hat{\bf{k}}^{\prime} = {\bf{k}}^{\prime} / | {\bf{k}}^{\prime} | $ are
unit vectors.

The boson representations of the spin-spin correlation functions are obtained in a
similar manner. To calculate the dynamic structure factor at finite temperatures,
we use the fluctuation dissipation theorem to express the dynamic structure factor
in terms of the corresponding imaginary frequency spin  Green's function,
 \begin{equation}
 S ( {\bf{q}} , \omega )  =   \left[ 1 + n ( \hbar \omega / T ) \right]
 \frac{ 2 \hbar}{T} {\rm Im} \left[ {\cal{G}} ( {\bf{q}} , i \omega_n )
 \right]_{ i \omega_n \rightarrow \omega + i 0^{+} }
 \label{eq:fluc1}
 \; \; \; ,
 \end{equation}
where 
$n (x ) = [ e^x -1 ]^{-1}$ is the Bose-Einstein occupation factor, and
 \begin{equation}
 {\cal{G}} ( {\bf{q}} , i \omega_n ) =
 \sum_{\bf{r}} T \int_0^{1/T} d \tau e^{ - i ( {\bf{q}} \cdot {\bf{r}} - \omega_n \tau )}
 \frac{1}{Z} Tr \left\{ e^{- \hat{H}/T} {\bf{S}}_{\bf{r}} ( \tau ) \cdot {\bf{S}}_0 ( 0 ) 
 \right\}
 \; \; \; .
 \label{eq:spingreens}
 \end{equation}
Here ${\bf{S}}_{\bf{r}} ( \tau ) = e^{\hat{H} \tau } {\bf{S}}_{\bf{r}} ( 0 )
e^{- \hat{H} \tau }$, and $\omega_n = 2 \pi n T $ are bosonic  Matsubara
frequencies.  The relation between the staggered structure factor
$S_{\rm st} ( {\bf{q}} , \omega )$ and the corresponding spin Green's function
${\cal{G}}_{\rm st} ( {\bf{q}} , i \omega_n )$ is identical with Eq.(\ref{eq:fluc1}). The
definition  of the staggered spin Green's function can be obtained 
from Eq.(\ref{eq:spingreens})
by inserting an additional factor of $e^{ i{\bf{\Pi}} \cdot {\bf{r}}}$ in the sum.
The spin Green's functions ${\cal{G}} ( {\bf{q}}, i \omega_n )$ and
${\cal{G}}_{\rm st} ( {\bf{q}}, i \omega_n )$ can be calculated via 
a conventional time-ordered perturbation expansion within the Matsubara
formalism for bosonic many-body systems.

\section{The leading behavior of $1/T_1^{\rm O}$}
\label{sec:oneloop}

According to Eqs.(\ref{eq:NMRdynstruc2}--\ref{eq:fstO1}) the 
$^{17}$O NMR-rate is at low temperatures given by 
 \begin{equation}
 \frac{1}{T_1^{\rm O}} = 
 \frac{  A_{\rm O}^2}{ \hbar^2 } \left[ \frac{2}{3} F (T )  + \frac{1}{3} F_{\rm st} ( T ) \right]
 \label{eq:NMRO}
 \; \; \; ,
 \end{equation}
where 
 \begin{eqnarray}
 F ( T ) & = & \frac{2}{N} {\sum_{\bf{q}}}^{\prime} S ( {\bf{q}} , 0 ) 
 \label{eq:Fdef}
 \; \; \; ,
 \\
 F_{\rm st} ( T ) & = & \frac{2}{N} {\sum_{\bf{q}}}^{\prime} \frac{ ( q_x a )^2 }{2}
 S_{\rm st} ( {\bf{q}} , 0 ) 
 \label{eq:Fstdef}
 \; \; \; .
 \end{eqnarray}
From Eqs.(\ref{eq:DMa}) and (\ref{eq:DMb}) it is clear that
within the Dyson-Maleev  transformation one obtains in general
two distinct contributions to the dynamic structure factor,
 \begin{equation}
 S ( {\bf{q}} , \omega ) = S_1 ( {\bf{q}}, \omega ) + S_2 ( {\bf{q}} , \omega )
 \label{eq:S12}
 \; \; \; ,
 \end{equation}
(and analogously for $S_{\rm st } ( {\bf{q}} , \omega )$), where
the first term $S_1 ( {\bf{q}} , \omega )$  is due to  
the one-magnon part of the
spin Green's function ${\cal{G}} ( {\bf{q}} , i \omega_n )$, while
$S_2 ( {\bf{q}} , \omega )$  is due to the two-magnon part (i.e. it 
involves a product of two magnon creation and two
annihilation operators). Accordingly, we shall refer to $S_1 ( {\bf{q}} , \omega )$ 
as the {\it{one-magnon part}}, and
to $S_2 ( {\bf{q}} , \omega )$ as  the {\it{two-magnon part}} 
of the dynamic structure factor.
Note that the averages are defined with respect to the interacting
magnon Hamiltonian, so that both terms involve also higher order virtual magnon excitations.
However, at the one-loop order it is sufficient to evaluate the expectation values
with the free magnon Hamiltonian $\hat{H}^{(2)}_{\rm DM}$. 
In this approximation the one-magnon term $S_1 ( {\bf{q}} , \omega )$
as well as the transverse part $S_{2 , \bot } ( {\bf{q}} , \omega )$
of the two-magnon contribution
vanish in the limit $\omega \rightarrow 0$, so that
these terms do not contribute to Eq.(\ref{eq:Fdef}). 
After a simple calculation we thus obtain to leading order 
 \begin{equation}
 F ( T ) \approx F^{(1)} ( T) = \frac{ 2 \hbar a^4 }{\pi} \int_{0}^{\infty} dq q \int_0^{\infty} d k k
 n ( \lambda_{\rm th} k ) [ 1 + n ( \lambda_{\rm th} q ) ]
 \delta ( \hbar c ( k - q ))
 \label{eq:Fres1}
 \; \; \; ,
 \end{equation}
where the superscript $(1)$ indicates the one-loop approximation.
We have written
$E_{\bf{k}} / T =  \lambda_{\rm th}| {\bf{k}} | $,
so that it is obvious that 
the Bose-Einstein factors act as cutoff functions which eliminate
the contribution  from modes with wavelengths short compared with the
thermal de Broglie wavelength $\lambda_{\rm th}$. 
Using the fact that by symmetry we may replace $q_x^2 \rightarrow {\bf{q}}^2 /2$
in Eq.(\ref{eq:Fstdef}), we obtain for the corresponding staggered function
$F_{\rm st} ( T )$ {\it{exactly the same expression}} as in Eq.(\ref{eq:Fres1}).
Scaling out the temperature dependence by defining $x = q \lambda_{\rm th}$,
we obtain to leading order
 \begin{equation}
 F^{(1)} ( T ) = F_{\rm st}^{(1) } ( T ) = 
 \frac{2 \pi}{3} \frac{a}{c} \left( \frac{a}{\lambda_{\rm th}}
 \right)^3 =
 \frac{2 \pi}{3} \frac{a}{c} 
 \left( \frac{Ta}{\hbar c } \right)^3 
 \label{eq:F1res}
 \; \; \; .
 \end{equation}
Substituting this into Eq.(\ref{eq:NMRO}), we finally obtain\cite{footnote2}
 \begin{equation}
 \frac{1}{T_1^{\rm O}} =  C_1 \frac{  A_{\rm O}^2  }{  \hbar^2}  \frac{a}{c}
 \left( \frac{Ta}{\hbar c } \right)^3 
 \; \; \; , \; \; \; C_{1} = \frac{2 \pi}{3} \approx 2.09
 \; \; \; .
 \label{eq:T1Ores}
 \end{equation}
We would like to emphasize two points:  First of all,
from Eq.(\ref{eq:Fres1}) it is obvious that the $T^3$-behavior of
$1/T_1^{\rm O}$ is a simple consequence of the two-dimensional phase space
and the energy conservation of antiferromagnetic magnons with linear energy
dispersion. The second important point is that the numerical value of the
prefactor in Eq.(\ref{eq:T1Ores}) is determined by non-critical fluctuations 
of the total magnetization and by fluctuations of the staggered magnetization
with typical wavelengths of the order of $1 / \lambda_{\rm th}$. 
From  Eqs.(\ref{eq:NMRO}) and (\ref{eq:Fres1}) it is clear
that to leading order the relative weight of the
total magnetization fluctuations is exactly twice as large as the weight of the
staggered fluctuations.

The experimental confirmation of Eq.(\ref{eq:T1Ores}) would give direct evidence
of the existence of short wavelength magnons in two-dimensional Heisenberg
antiferromagnets at low temperatures.
In the recent experiment by Thurber {\it{et al.}}\cite{Thurber96}
the $^{17}$O-NMR rate in  
Sr$_2$CuO$_2$Cl$_2$ 
was measured with sufficient accuracy to
confirm the $T^{3}$-behavior above the Neel temperature,
and to obtain an estimate
for the numerical value for the prefactor  $C_{1}$. 
The experimental
result is\cite{Thurber96} $C_{1} \approx 2.7 \pm 0.9 $, where the experimental uncertainty
of $30 \%$ is mainly due to the ambiguities in the measurements of the
hyperfine-coupling $A_{\rm O}$ and the exchange coupling $J$\cite{Imaipriv}.
We conclude that within the experimental accuracy the lowest order
spin-wave result agrees with the experiment\cite{Thurber96}.
This is a very important result, because it demonstrates that,
to a very good approximation,
short wavelength spin-waves in two-dimensional Heisenberg magnets
can be treated as free particles.
We would like to emphasize that 
there is no long-range order in the system,
so that well-defined magnons
in two-dimensional Heisenberg magnets at finite temperatures 
reflect fundamentally different physics
than magnons in three-dimensional magnets below the Neel temperature.
In fact, in two dimensions and at $T > 0$
magnons cannot propagate over length scales larger than
the correlation length\cite{CHN88}
$\xi \propto \exp [ 2 \pi \rho_s / T]$, where $\rho_s$ is the spin stiffness.
In other words, at short
distances the interactions between the magnons become weak, while
at long distances the magnons completely disappear from the physical spectrum.

\section{The two-loop correction}
\label{sec:twoloop}

As noticed in Ref.\cite{Chacky91},
the two-loop correction to
Eq.(\ref{eq:T1Ores}) yields a contribution proportional to $T^4$, 
which for an ideal two-dimensional system is
negligible at sufficiently low temperatures. Although in the appendix of
Ref.\cite{Chacky91} it was shown how the two-loop correction
can be obtained in principle, the numerical value of the coefficient 
of the $T^4$-term was not calculated, because a few years ago 
sufficiently accurate experimental data to test the theoretical prediction were not available.
Thurber {\it{et al.}} \cite{Thurber96}  have reported for the first time high-quality data
of $1/T_1^{\rm O}$ in an almost ideal two-dimensional quantum Heisenberg
antiferromagnet, which are accurate enough to allow for a detailed
comparison with theory.
Motivated by this experiment, we shall now calculate the numerical value of the
two-loop correction.

In quantum antiferromagnets the evaluation higher-order corrections to
the leading terms in the spin wave-expansion  is a rather difficult task,
because after the Bogoliubov transformation (\ref{eq:Bogoliubov})
the two-body part $\hat{H}^{(4)}_{\rm DM}$ of the spin-wave 
Hamiltonian (see Eq.(\ref{eq:HDM4})) and the boson representation of the
spin operators involve many different terms.
Although the one-loop calculation can be performed 
in a straight-forward manner without resorting to
elaborate many-body techniques, 
the classification of the
large number of terms contributing to the two-loop correction
is greatly facilitated with the help of
a graphical representation in terms of
Feynman  diagrams\cite{Kopietzphd}.

\subsection{The two-magnon part of the total spin structure factor}
\label{subsec:Stot2mag}

We begin with the calculation of the two-loop correction to the two-magnon part
$S_2 ( {\bf{q}} , 0 )$ of the total spin structure factor.
This is obtained by contracting the interaction part $\hat{H}^{(4)}_{\rm DM}$ 
with the two-magnon part of the spin-spin correlation function. 
Diagrammatically the longitudinal and the transverse components of the
dynamic structure factor are calculated separately,
 \begin{equation}
 S_2^{(2)} ( {\bf{q}}, 0 ) = 
 S_{2zz}^{(2)} ( {\bf{q}}, 0 ) + 
 S_{2 \bot}^{(2)} ( {\bf{q}}, 0 ) 
 \label{eq:S2decomp}
 \; \; \; .
 \end{equation}
The superscript $(2)$ denotes the two-loop approximation,
while the subscripts $2zz$ and $2 \bot$ denote the
longitudinal and transverse part of the
two-magnon term in the dynamic structure factor.
Denoting by
$F^{(2)} ( T )= F_{\bot}^{(2)} ( T ) + F_{zz}^{(2)} ( T )$
the corresponding two-loop correction
to the dimensionless function $F ( T )$ defined in Eq.(\ref{eq:Fdef}),
we have at two-loop order
 \begin{equation}
 F ( T ) \approx  F^{(1)} ( T ) + F^{(2)}  ( T )
 \label{eq:Ftwoloop}
 \; \; \; ,
 \end{equation}
where the one-loop contribution $F^{(1)} ( T )$ is given in
Eq.(\ref{eq:F1res}).
Using the fact that in two dimensions and for large $S$ the spin-wave velocity
and the spin-stiffness are related via
$\hbar c = {2 \sqrt{2} \rho_s a }/{S}$,
we find after some tedious algebra\cite{Kopietzphd} 
that the transverse part can be written as
 \begin{equation}
 F_{\bot}^{(2)} ( T ) = \frac{4}{\pi} \frac{a}{c} \left( \frac{T a}{\hbar c} \right)^3
 \left( \frac{T}{2 \pi \rho_s} \right)
 \left[ K_1 +
 \int_0^{\infty} dx \int_0^{\infty} dy x^2 n ( x ) n (y) [ 1 + n ( x+y)]
 \right]
 \label{eq:F2botres}
 \; \; \; ,
 \end{equation}
and the longitudinal part is
 \begin{equation}
 F_{zz}^{(2)} ( T )  =   \frac{4}{\pi} \frac{a}{c} \left( \frac{T a}{\hbar c} \right)^3
 \left( \frac{T}{2 \pi \rho_s} \right) \left[ -K_2 - K_3 -
 \int_0^{\infty} dx \int_0^{\infty} dy x^2 n ( x ) n (y) [ 1 + n ( x)]
 \right]
 \; \; \; .
 \label{eq:F2zzint1}
 \end{equation}
Here $K_1$, $K_2$ and $K_3$ are numerical constants of the order
of unity, which are given in the Appendix.
As already mentioned in Sec.\ref{sec:DM}, 
zero temperature $1/S$-renormalizations are
implicitly taken into account in our final result (\ref{eq:TOres})
by identifying $c$ and $\rho_s$ with the experimentally 
measured parameters.
The last integrals in the square braces of Eqs.(\ref{eq:F2botres}) 
and (\ref{eq:F2zzint1}) are logarithmically
divergent. This is not surprising, because $F_{\bot}^{(2)} ( T )$
and $F_{zz}^{(2)} ( T )$ 
are not rotationally invariant, and are therefore
sensitive to the fact that
our two-dimensional spin system does not have long-range order at 
any $T > 0$.  As discussed at the end of Sec.\ref{sec:intro}, 
the logarithmic divergence should cancel in 
the rotationally invariant function
$F_{\bot}^{(2)}  ( T ) + F_{zz}^{(2)} (T)$. 
Combining Eqs.(\ref{eq:F2botres}) and (\ref{eq:F2zzint1}), 
it is easy to see that this is indeed the case, and
we finally obtain
 \begin{equation}
 F^{(2)} ( T )  
 =
 \frac{4}{\pi} \frac{a}{c} \left( \frac{T a}{\hbar c} \right)^3
 \left( \frac{T}{2 \pi \rho_s} \right)
 \left[ {K_1} -  K_2 - K_3 -  K_4 \right]
 \label{eq:F2res}
 \; \; \; ,
 \end{equation}
where the {\it{finite}} numerical constant $K_4$ is 
given in Eq.(\ref{eq:K4def}).
Comparing Eq.(\ref{eq:F2res}) with Eq.(\ref{eq:Fres1}), we conclude that
the two-loop correction involves an additional factor of
$T / ( 2 \pi \rho_s )$, so that it is indeed negligible
at sufficiently low temperatures.
However, as discussed below, for a quantitative comparison with the 
experiment\cite{Thurber96} this correction can be substantial.

\subsection{The two magnon part of the staggered structure factor}

According to Eq.(\ref{eq:NMRO}), the rate $1/T_1^{\rm O}$ 
receives also a contribution from
the weighted average $F_{\rm st}  (T )$ of the staggered structure factor
$S_{\rm st} ( {\bf{q}} , 0)$ defined in Eq.(\ref{eq:Fstdef}).
Let us first calculate the two-magnon part $S_{{\rm st},2} ( {\bf{q}} , \omega )$.
Similar to Eq.(\ref{eq:S2decomp}), we obtain
at two-loop order
 \begin{equation}
 S_{{\rm st},2}^{(2)} ( {\bf{q}}, 0 ) = 
 S_{{\rm st},2zz}^{(2)} ( {\bf{q}}, 0 ) + 
 S_{{\rm st}, 2 \bot}^{(2)} ( {\bf{q}}, 0 ) 
 \label{eq:Sst2decomp}
 \; \; \; .
 \end{equation}
The corresponding contributions
$F_{{\rm st}, 2 \bot}^{(2)} ( T )$ and
$F_{{\rm st}, 2 zz}^{(2)} ( T )$ 
to the function $F_{\rm st} ( T )$, are
 \begin{equation}
 F^{(2)}_{{\rm st},2  \bot} ( T) =
 F^{ (2) }_{\bot} ( T)
 \; \; \; ,
 \label{eq:F2botequal}
 \end{equation}
 \begin{equation}
 F_{{\rm st},2zz}^{(2)} ( T )   =   
 \frac{4}{\pi} \frac{a}{c} \left( \frac{T a}{\hbar c} \right)^3
 \left( \frac{T}{2 \pi \rho_s} \right) \left[ -K_5   - J \right]
 \; \; \; ,
 \label{eq:Fst2zzint1}
 \end{equation}
where the numerical constant $K_5$ 
is given in Eq.(\ref{eq:K5def}), and the divergent integral $J$ is given in
Eq.(\ref{eq:Jdef}).
Combining Eqs.(\ref{eq:F2botequal}) and (\ref{eq:Fst2zzint1}), we obtain
 \begin{equation}
 F_{{\rm st},2}^{ (2)} ( T ) \equiv F_{{\rm st},2 \bot}^{(2)} ( T ) + 
 F_{{\rm st},2zz}^{ (2)} ( T ) 
 =
 \frac{4}{\pi} \frac{a}{c} \left( \frac{T a}{\hbar c} \right)^3
 \left( \frac{T}{2 \pi \rho_s} \right)
 \left[ {K_1} -  K_5 - K_6 \right]
 \label{eq:F2stres}
 \; \; \; ,
 \end{equation}
where the {\it{finite}} numerical constant $K_6$ is given in Eq.(\ref{eq:K6def}).

\subsection{The one-magnon part of the staggered structure factor}

There exists one more contribution to the function
$F_{\rm st}^{(2)} ( T )$, which is due fact that at two-loop order
the magnon self-energy acquires a finite imaginary part.
This generates a finite
one-magnon contribution
$S_{{\rm st},1} ( {\bf{q}} , 0 )$ 
to the staggered structure factor, which in turn gives rise
to an additional two-loop contribution
$F_{{\rm st},1}^{(2)} ( T )$ to $F_{\rm st} ( T )$. This contribution should be added
to Eq.(\ref{eq:F2stres}) in order to obtain the total two-loop correction.
Hence, at two-loop order we have
(compare with Eq.(\ref{eq:Ftwoloop}))
 \begin{equation}
 F_{{\rm st}} ( T ) \approx F_{{\rm st}}^{(1)} ( T ) + F_{{\rm st},2 }^{(2)} ( T ) + 
 F_{{\rm st},1}^{(2)} ( T )
 \label{eq:Fsttwoloop}
 \; \; \; .
 \end{equation}
By symmetry, no such one-magnon contribution arises in the
corresponding total spin structure factor.
To see this,
let us follow Harris {\it{et al.}}\cite{Harris71} and  denote by
$\Sigma_{\alpha \alpha } ( {\bf{q}} , \omega + i 0^{+} )$
the usual retarded self-energy for $\alpha$-magnons, and  by
$\Sigma_{\alpha \beta } ( {\bf{q}} , \omega + i 0^{+} )$
the off diagonal magnon self-energy.
The magnon damping can then be written as
 $\hbar \Gamma_{\mu \nu } ( {\bf{q}}, \omega ) 
 = - 2 {\rm Im} \Sigma_{\mu \nu } ( {\bf{q}} , \omega + i 0^{+} )$,
where $\mu ,\nu = \alpha , \beta$. 
The one-magnon contribution to the 
total and staggered dynamic structure factor at vanishing
frequency is\cite{Harris71,Kopietzphd}
 \begin{eqnarray}
 S_1 ( {\bf{q}} , 0 )  & = & 2 \hbar S  \frac{ u_{\bf{q}}^2 (1 - x_{\bf{q}})^2 }{ E_{\bf{q}}^2}
 \lim_{\omega \rightarrow 0} \left[
 \hbar \tilde{\Gamma}_{\alpha \alpha} ( {\bf{q}} , \omega ) + \hbar
 \tilde{\Gamma}_{\alpha \beta } ( {\bf{q}} , \omega ) \right]
 \label{eq:S1gamma}
 \; \; \; ,
 \\
 S_{{\rm st},1} ( {\bf{q}} , 0 )  & = & 2 \hbar S 
  \frac{ u_{\bf{q}}^2 (1 + x_{\bf{q}})^2 }{ E_{\bf{q}}^2}
 \lim_{\omega \rightarrow 0} \left[
 \hbar \tilde{\Gamma}_{\alpha \alpha} ( {\bf{q}} , \omega ) - \hbar
 \tilde{\Gamma}_{\alpha \beta } ( {\bf{q}} , \omega ) 
  \right]
 \label{eq:S1stgamma}
 \; \; \; ,
 \end{eqnarray}
where we have defined 
 \begin{equation}
 \tilde{\Gamma}_{\mu \nu } ( {\bf{q}} , \omega ) =
 \frac{ \Gamma_{\mu \nu } ( {\bf{q}} , \omega )}{
  1 - e^{- \hbar \omega / T} }
  \; \; \; , \; \; \; \mu , \nu = \alpha , \beta
  \label{eq:tildegamma}
  \; \; \; .
  \end{equation}
Noting that\cite{Harris71} 
$\Gamma_{\alpha \alpha} ( {\bf{q}} , \omega ) = 
\Gamma_{\alpha \beta} ( {\bf{q}} , - \omega )$, we see that
for $\omega \rightarrow 0$ the one-magnon contribution 
to the total spin structure factor vanishes by symmetry,
$S_1 ( {\bf{q}} , 0 ) = 0$.
On the other hand, the one-magnon contribution to the staggered structure factor
has a finite contribution at vanishing frequency, which can be written as
 \begin{equation}
 S_{{\rm st},1} ( {\bf{q}} , 0 )   =  4 \hbar S  
 \frac{ u_{\bf{q}}^2 (1 + x_{\bf{q}})^2 }{ E_{\bf{q}}^2}
 \lim_{\omega \rightarrow 0} 
 \hbar \tilde{\Gamma}_{\alpha \alpha} ( {\bf{q}} , \omega ) 
 \label{eq:S1stfinite}
 \; \; \; .
 \end{equation}
Calculating the (rescaled) damping rate
$\tilde{\Gamma}_{\alpha \alpha} ( {\bf{q}} , 0 )$ at two-loop 
order\cite{Harris71,Kopietzphd}, we obtain after some tedious algebra
 \begin{equation}
 F_{{\rm st},1}^{(2)} ( T )   =   
 \frac{4}{\pi} \frac{a}{c} \left( \frac{T a}{\hbar c} \right)^3
 \left( \frac{T}{2 \pi \rho_s} \right) K_7
 \; \; \; ,
 \label{eq:Fst2zzintres}
 \end{equation}
with the numerical constant $K_7$ given in Eq.(\ref{eq:K7}).

\subsection{The final result}

Combining the results of the previous three subsections
with  the one-loop result given in
Eq.(\ref{eq:F1res}), we conclude that at low temperatures
the two-loop corrections to the functions $F ( T )$ and $F_{\rm st} ( T )$ can
be written as
 \begin{eqnarray}
 F^{(2)} ( T  ) & = & K \left( \frac{ T}{ 2 \pi \rho_s } \right) F^{(1)} ( T ) 
 \label{eq:F21}
 \; \; \; ,
 \\
 F_{\rm st}^{(2)} ( T  ) & = & K_{\rm st} \left( \frac{ T}{ 2 \pi \rho_s } 
 \right) F^{(1)}_{\rm st} ( T ) 
 \label{eq:Fst21}
 \; \; \; ,
 \end{eqnarray}
with
 \begin{eqnarray}
 K & = & \frac{6}{\pi^2} \left[ {K_1} - K_2 - K_3 - K_4 \right]
 \approx -2.19
 \label{eq:Kres}
 \; \; \; ,
 \\
 K_{\rm st} & = & \frac{6}{\pi^2} \left[ K_1 - K_5 - K_6 + K_7 \right]
 \approx -1.27
 \label{eq:Kstres}
 \; \; \; .
 \end{eqnarray}
From Eq.(\ref{eq:NMRO}) we thus conclude that at low-temperatures the
$^{17}$O NMR rate is given by
 \begin{equation}
 \frac{1}{T_1^{\rm O}} =   \frac{ 2 \pi  A_{\rm O}^2  }{3  \hbar^2}  \frac{a}{c}
 \left( \frac{Ta}{\hbar c } \right)^3 \left[ 1 + C_2
 \left( \frac{T}{2 \pi \rho_s} \right)
 + O ( T^2 ) \right]
 \; \; \; , 
 \label{eq:TOres}
 \end{equation}
with
 \begin{equation}
 C_2 = \frac{2}{3} K + \frac{1}{3} K_{\rm st} \approx -1.88
 \; \; \; .
 \label{eq:C2res}
 \end{equation}
This is our main result. 
Note that the energy scale of the two-loop correction
is set by the spin-stiffness $\rho_s$, 
so that a measurement of the
sub-leading correction can in principle be used to obtain the spin stiffness
of the material.
Because the  coefficient $C_2$ is
negative, the two-loop correction leads to a reduction of the NMR-rate.
In other words: Non-interacting spin-wave theory over-estimates the
magnitude of the NMR-rate of oxygen.
Although this reduction becomes negligible at sufficiently low temperatures,
it is substantial in the experiment of Thurber {\it{et al.}}\cite{Thurber96}.
A simple estimate\cite{Imaipriv} for the spin stiffness in this experiment
yields
$ 2 \pi \rho_s \approx 1500 K$. Taking into account that
$T \approx 300 K$ in the low-temperature regime above the Neel temperature, 
we conclude that the correction in
Eq.(\ref{eq:TOres})  leads to a {\it{reduction}} of the one-loop result by
almost $40 \%$. Because the measured rate was approximately $30 \%$ {\it{larger}} than
the theoretical one-loop result, we conclude that the
agreement between theory and experiment becomes worse if the two-loop correction
is included.

\section{Summary and conclusions}

Motivated by recent measurements 
of the NMR-relaxation rates in the quasi-two-dimensional
antiferromagnet Sr$_2$CuO$_2$Cl$_2$ by
Thurber {\it{et al.}}\cite{Thurber96},
we have calculated the
$^{17}$O NMR-rate above the Neel temperature with the help of the spin-wave
expansion at two-loop order. If magnon-magnon interactions are ignored, a
calculation based on the two-dimensional isotropic Heisenberg antiferromagnet 
leads to a quantitative agreement with the  data
within the experimental accuracy. 
We have pointed out that $1/T_1^{\rm O}$ is dominated by magnons
with wavelengths of the  order of the thermal de Broglie wavelength
$\lambda_{\rm th}$, so that the agreement between lowest order spin-wave theory and
experiment shows that short wavelength magnons are 
indeed well-defined elementary excitations in 
two-dimensional Heisenberg magnets at low temperatures.
We would like to emphasize that the existence of finite temperature
magnons in two dimensions reflects fundamentally different physics than
in three-dimensional systems  below the Neel-temperature.

We have also calculated  the leading correction to the 
$T^{3}$-behavior due to magnon-magnon interactions. 
This correction
involves an additional power of $T / ( 2 \pi \rho_s )$, 
i.e. the energy scale for the correction is set 
by the spin stiffness $\rho_s$.
The numerical value of the prefactor of the $T^4$-term
is calculated here for the first time.
Our main result is given in Eqs.(\ref{eq:TOres}) and (\ref{eq:C2res}), 
and implies that the quantitative agreement between 
spin-wave theory and the experiment 
becomes less convincing if
magnon-magnon interactions are taken into account.
It should be kept in mind, however, that 
we have modeled the system by an isotropic 
two-dimensional Heisenberg antiferromagnet.
In the low temperature regime close to the Neel temperature
the various  anisotropies and the finite interchain coupling 
in the experimental system will certainly become important,
so that Eq.(\ref{eq:TOres}) cannot be trusted for temperatures too close to
the Neel temperature. 
This might partially explain the slight discrepancy 
between theory and experiment.
Another potentially important contribution
to the NMR rate, which is not contained in our perturbative spin-wave calculation, is 
due to spin-diffusion\cite{Chakravarty90}.
However, this contribution should be suppressed by anisotropies.
Very recently, a measurement of the frequency-dependence of
$1/T_1^{\rm O}$ in Sr$_2$CuO$_2$Cl$_2$ 
showed that the contribution from spin diffusion 
is indeed negligible\cite{Imaipriv}.

\section*{Acknowledgments}
We would like to thank Takashi Imai for convincing us that 
for the interpretation of the experiment\cite{Thurber96}
it is important to know the precise numerical value 
of the $T^4$ correction to the $^{17}$O NMR-rate, and 
for explaining to us some experimental details.
We would also like Kent Thurber for his comments on 
the manuscript.
The work of S. C. was supported by the National Science Foundation,
Grant No. DMR-9531575.

\appendix

\section*{}
In this appendix we give the expressions  for the 
numerical constants $K_i$, $i = 1, \ldots , 7$ 
introduced in Sec.\ref{sec:twoloop}.
The numerical constant $K_1$ in Eq.(\ref{eq:F2botres}) is given by
 \begin{equation}
 K_1 = \frac{1}{2}
 \int_0^{\infty} dx \int_0^{\infty} dy x y n ( x ) n (y) [ 1 + n ( x+y)]
 \approx 1.80
 \; \; \; .
 \label{eq:K1def}
 \end{equation}
The constants $K_2$ and $K_3$ that appear in 
Eq.(\ref{eq:F2zzint1}) can be written as the following four-dimensional
integrals,
 \begin{eqnarray}
 K_2 & = &   \frac{1}{\pi^2}
 \int d^{2} {\bf{x}}  \int d^{2} {\bf{y}} \;  x y  n ( x ) n (y) [ 1 + n ( x)]
\frac{  1 }{
| \hat{\bf{q}} x - {\bf{x}} - {\bf{y}} |^2 - y^2 }
\label{eq:K2def}
\; \; \; ,
\\
K_3 & = & 
  \frac{1}{8 \pi^2} \int d^{2} {\bf{x}} \int d^{2} {\bf{y}} \; \frac{x}{y} 
 n (x) [ 1 + n (x) ]
\frac{  | \hat{\bf{q}} x - {\bf{x}} - {\bf{y}} | - y }{
| \hat{\bf{q}} x - {\bf{x}} - {\bf{y}} | + y }
 \; \; \; .
 \label{eq:K3def}
 \end{eqnarray}
Here $\hat{\bf{q}}$ is an arbitrary fixed unit vector.
The integral (\ref{eq:K2def}) can be
reduced to a two-dimensional one with the help of 
circular coordinates in the
${\bf{x}}$-  and ${\bf{y}}$-planes, which is then easily
calculated numerically,
 \begin{equation}
 K_2  = \int_0^{\infty} dx \int_0^{x} dy \; x y n (x ) n (y) [ 1 + n ( x) ]
 \approx 2.09
 \label{eq:K2res}
 \; \; \; .
 \end{equation}
The integral $K_3$ can be calculated analytically 
by shifting
${\bf{y}} \rightarrow {\bf{y}} - {\bf{r}}/2$, with
${\bf{r}} = {\bf{x}} - x \hat{\bf{q}}$, and 
then introducing circular coordinates $(x, \varphi)$ in the ${\bf{x}}$-plane,
and elliptic coordinates $( u , \alpha)$
in the ${\bf{y}}$-plane, i.e.
${\bf{y}} \cdot \hat{\bf{r}}  =  \frac{r}{2} \cosh u \cos \alpha$, and
${\bf{y}} \cdot \hat{\bf{r}}_{\bot}  =  \frac{r}{2} \sinh u \sin \alpha$.
Here $r = | {\bf{r}} | = x \sqrt{ 2 ( 1 - \hat{\bf{q}} \cdot {\hat{\bf{x}}} )}$, and
${\hat{\bf{r}}}_{\bot}$ is a unit vector
perpendicular to ${\hat{\bf{r}}}$.
With these coordinates
the integrations in $K_3$ factorize and can then be done exactly, with the result
 \begin{equation}
 K_3 = \frac{3}{2} \zeta ( 3 ) \approx 1.80
 \label{eq:K3res}
 \; \; \; .
 \end{equation}
The constant $K_4$ in Eq.(\ref{eq:F2res}) is given by
 \begin{equation}
 K_4 =
 \int_0^{\infty} dx \int_0^{\infty} dy x^2 n ( x ) n (y) [ n(x)  - n ( x+y)]
 \approx 1.52
 \label{eq:K4def}
 \; \; \; .
 \end{equation}
Note that the first term in the square brace of Eq.(\ref{eq:K4def})
is due to longitudinal fluctuations, while the second term is
generated by transverse fluctuations.
Each of these terms separately would lead to a logarithmically divergent integral,
but the divergence cancels in the rotationally invariant 
function $F^{(2)} ( T )$, in agreement with Ref.\cite{David81}.
This is also a non-trivial check that in our two-loop calculation we did not
forget any diagram.

The numerical constant $K_5$ in Eq.(\ref{eq:Fst2zzint1}) is
 \begin{equation}
 K_5 =
 - \frac{1}{4 \pi^2} \int d^{2} {\bf{x}} \int d^{2} {\bf{y}}
 \frac{x^2}{y} n ( x ) [ 1 + n ( x) ]
 \frac{ (1 - \hat{\bf{x}} \cdot \hat{\bf{q}} )
\hat{\bf{x}} \cdot ( x \hat{\bf{q}}  - {\bf{x}} - {\bf{y}} ) }{ 
( | x \hat{\bf{q}} - 
{\bf{x}} - {\bf{y}} | + y ) 
| x \hat{\bf{q}}  - {\bf{x}} - {\bf{y}} | }
\label{eq:K5def}
\; \; \; .
\end{equation}
This is a constant of the order of unity,
which can be calculated analytically
with the help of circular coordinates in the
${\bf{x}}$-plane and (after shifting 
${\bf{y}} \rightarrow {\bf{y}} - \frac{1}{2} ( {\bf{x}} - x \hat{\bf{q}} ) $)
elliptic coordinates in the ${\bf{y}}$-plane. 
We obtain
 \begin{equation}
 K_5 = 4 \zeta ( 3 ) \approx 4.80
 \; \; \; .
 \label{eq:K5res}
 \end{equation}
The integral $J$ in Eq.(\ref{eq:Fst2zzint1}) is
 \begin{equation}
 J =
  \frac{1}{2 \pi^2} \int d^{2} {\bf{x}} \int d^{2} {\bf{y}}
 \frac{x^3}{y} n ( x ) [ 1 + n ( x) ] n (y)
 \frac{ (1 - \hat{\bf{x}} \cdot \hat{\bf{q}} )^2 }{
| x \hat{\bf{q}} - 
{\bf{x}} - {\bf{y}} |^2 - y^2   }
\label{eq:Jdef}
\; \; \; .
\end{equation}
Introducing circular coordinates in the ${\bf{x}}$- and ${\bf{y}}$-planes,
the two angular integrations can be performed analytically,
with the result
 \begin{equation}
 J = \int_0^{\infty} dx \int_0^{x} dy ( x^2 + y^2 ) n ( x ) [1 + n (x) ] n (y )
 \label{eq:Jres}
 \; \; \; .
 \end{equation}
Note that for small $y$ the
Bose-Einstein factor $n(y)$
diverges as $1/y$, leading to a
logarithmic divergence of the integral $J$.
However, in the rotationally invariant correlation function this
divergence is again cancelled, so that the constant $K_6$
in Eq.(\ref{eq:F2stres}) is finite,
 \begin{equation}
 K_6 =
 \int_0^{\infty} dx \int_0^{x} dy (x^2 + y^2) n ( x ) n (y) [ n(x)  - n ( x+y)]
 \approx 0.89
 \label{eq:K6def}
 \; \; \; .
 \end{equation}
Finally,
the numerical constant $K_7$ in Eq.(\ref{eq:Fst2zzintres})
is given by
 \begin{eqnarray}
 K_7 & = & \frac{1}{2} \int_0^{\infty} dx \int_0^{\infty} dy x^2 n ( x ) n (y ) [ 1 + n (x + y )]
 \left[ 1 + \frac{y}{x + y } \right]
 \left[ 1 - \frac{x}{\sqrt{ x^2 + 4 y ( x + y ) } } \right]
 \nonumber
 \\
 & \approx & 1.80 
 \; \; \; .
 \label{eq:K7}
 \end{eqnarray}



\begin{thebibliography}{99}
\bibitem{footnote1}
The form factor given in Eq.(A3) of Ref.\cite{Chacky91} 
is too small by a factor of two; 
similarly, in Eq.(A4) of Ref.\cite{Chacky91} a factor
of four is missing.
\end{thebibliography}
\end{document}